\documentclass[12pt,amsmath,amssymb,preprint]{aastex}

\makeatother

\shorttitle{The Milky Way's dark matter halo}

\shortauthors{Saha et al.}

\begin{document}

\title{The Milky Way's dark matter halo appears to be lopsided}

\author{Kanak Saha.$^{1,2}$, Evan. S. Levine$^3$, Chanda J. Jog$^1$ and Leo Blitz$^3$}
\affil{$^{1}$Department of Physics, Indian Institute of Science, Bangalore 560012, India\\$^2$Academia Sinica Institute of Astronomy and Astrophysics - TIARA, P.O. Box 23-141, Taipei 10617, Taiwan \\$^3$Department of Astronomy, University of California at Berkeley, USA\\e-mail: kanak@physics.iisc.ernet.in}

\begin{abstract}
The atomic hydrogen gas (H {\footnotesize I}) disk in the outer region (beyond $\sim 10$ kpc from the centre) of Milky Way can provide valuable information about the structure of the dark matter halo. The recent 3-D thickness map of the outer H {\footnotesize I} disk from the all sky 21-cm line LAB survey, gives us a unique opportunity to investigate the structure of the dark matter halo of Milky Way in great detail. A striking feature of this new survey is the North-South asymmetry in the thickness map of the atomic hydrogen gas. Assuming vertical hydrostatic equilibrium under the total potential of the Galaxy, we derive the model thickness map of the H {\footnotesize I} gas. We show that simple axisymmetric halo models, such as softened isothermal halo (producing a flat rotation curve with $V_c \sim 220$ kms$^{-1}$) or any halo with density falling faster than the isothermal one, are not able to explain the observed radial variation of the gas thickness. We also show that such axisymmetric halos along with different H {\footnotesize I} velocity dispersion in the two halves, cannot explain the observed asymmetry in the thickness map. Amongst the non-axisymmetric models, it is shown that a purely lopsided ($m=1$, first harmonic) dark matter halo with reasonable H {\footnotesize I} velocity dispersion fails to explain the North-South asymmetry satisfactorily. However, we show that by superposing a second harmonic ($m=2$) out of phase onto a purely lopsided halo e.g. our best fit and more acceptable model A (with parameters $\epsilon_{h}^{1}=0.2$, $\epsilon_{h}^{2}=0.18$ and $\sigma_{HI}=8.5$ kms$^{-1}$) can provide an excellent fit to the observation and reproduce the North-South asymmetry naturally. The emerging picture of the asymmetric dark matter halo is supported by the $\Lambda$CDM halos formed in the cosmological N-body simulation.

\end{abstract}


\section{Introduction}
Asymmetries are common in disk galaxies and are seen in lopsidedness (Baldwin et al. 1980; Rix \& Zaritsky 1995; Bournaud et al. 2005; Saha et al. 2007), in warps (Garcia-Ruiz et al. 2002; Sanchez-Saavedra et al. 2003; Saha \& Jog 2006), in the rotation curves of receding side and approaching side (Manabe \& Miyamoto 1975; Swaters et al. 1999), and in the distribution of the neutral hydrogen gas in galaxies (Richter \& Sancisi 1994). These asymmetries in the dynamical phenomena can provide valuable information about the nature of the underlying dark matter potential. In fact, they have often been considered a reflection of the asymmetry in the dark matter distribution (Weinberg 1994; Jog 1997). However, one needs to be careful and systematically discard other possibilities such as dynamical instabilities as the origin of these asymmetries (e.g. Saha \& Jog 2006). Our motivation in this paper is to check whether a non-axisymmetric dark matter halo can explain the recently measured asymmetry in the atomic hydrogen gas distribution in our Galaxy. 

In order to study the shape of the dark matter halo, it is best to look for tracers which lie mostly outside the main baryonic disk component. Examples of such tracers are the motion of satellites, neutral hydrogen (H {\footnotesize I}) gas in the outer region of the galaxy etc.. In particular, over the last two decades it has become well known  that the H {\footnotesize I} gas layer flares significantly beyond the optical disk of our Galaxy (Kulkarni, Heiles \& Blitz 1982; Knapp 1987; Wouterloot et al. 1990; Diplas \& Savage 1991; Merrifield 1992; Nakanishi \& Sofue 2003) and the recent LAB (Leiden/Argentine/Bonn) survey (Kalberla et al. 2005) of Galactic H {\footnotesize I} reveals the most comprehensive, uniformly sampled flaring map of the H {\footnotesize I} gas extended out to a very large radius from the Galactic centre. Many external edge-on galaxies seem to show flaring in the thickness of the neutral hydrogen gas (Brinks \& Burton 1984; Olling 1996; Matthews \& Wood 2003). The most likely reason for the flaring is that the total gravitational force acting perpendicular to the disk plane decreases with radius while the velocity dispersion of H {\footnotesize I} is observed to be nearly constant (Lewis 1984). The contribution to the total perpendicular gravitational force comes mainly from the stellar disk, gas and the dark matter halo. Since the midplane density of the stars falls off rapidly compared to that of dark matter halo at large galactocentric radii (typically beyond the optical disk), the dark matter halo is expected to take over the major role in determining the vertical distribution of H {\footnotesize I} gas in the outer region.   
This makes the H {\footnotesize I} layer in the Galactic outskirts extremely sensitive to the distribution of dark matter and presently available neutral hydrogen gas (from the LAB survey) extended out to a very large radius from the Galactic centre provides us an unique opportunity to examine the detailed nature of the dark matter distribution in the Milky Way.

This approach has  been used to investigate the nature of dark matter halos by studying the thickness of the neutral hydrogen gas in our Galaxy (Olling \& Merrifield 1998, 2000, 2001, Narayan et al. 2005; Kalberla et al. 2007) and for M31 (Banerjee \& Jog 2008) . Of these, recent work by Narayan et al. 2005 based on the Wouterloot et al. (1990) H {\footnotesize I} data shows that the observed H {\footnotesize I} flaring of Milky Way is best fit by a spherical dark matter halo with density falling faster than the isothermal halo. On the other hand, the most recent and very detailed work by Kalberla et al. (2007) , based on the LAB survey data, shows that the dark matter distribution in the Galaxy is rather complicated. They needed a massive extended dark matter halo, a self-gravitating dark matter disk, and a dark matter ring to explain the flaring in the H {\footnotesize I} gas in our Galaxy. Now, one of the most striking feature in the LAB survey is that the neutral hydrogen gas thickness map shows systematic North-South (hereafter N-S) asymmetry, where Galactic North refers to $0\,^{\circ} < \varphi < 180\,^{\circ}$ and Galactic South by $180\,^{\circ} < \varphi < 360\,^{\circ}$, in the Galctocentric cylindrical polar coordinate system (R, $\varphi$, z). This asymmetry was previously seen in previous HI maps of the Milky Way (Henderson, Jackson and Kerr 1982), but rarely remarked on. So it may be worth investigating if the Milky Way H {\footnotesize I} flaring can be explained by a non-axisymmetric dark matter halo. 
 
Such non-axisymmetries in the collisionless dark matter halo are  probably not uncommon. In the current cosmological paradigm, the $\Lambda$ cold dark matter ($\Lambda$ CDM) halos are formed by dissipationless gravitational collapse of the material associated with the peaks of the primordial density fluctuation field and then grow via mergers and accretion in a highly nonlinear fashion. The resulting dynamical structures of these halos can be highly asymmetric. In fact, the $\Lambda$ CDM halos formed in the recent Millennium Simulation (Springel et al. 2005) show asymmetry in their mass distribution (Gao \& White 2006).     

In the present study, we derive, numerically, the flaring in the thickness of the neutral hydrogen gas using self-consistent model for the Galaxy (Narayan \& Jog 2002; Narayan et al. 2005) including a non-axisymmetric dark matter halo and disk. We take into account the self-gravity of the gas. Our analysis produces a non-axisymmetric flaring curve for the gas and in this respect our study is probably different from all the previous studies which tried to derive H {\footnotesize I} flaring in spiral galaxies. 
Based on our analysis, we show that an elliptically perturbed (readers are referred to \S 2.2 for details) lopsided dark matter halo with density falling faster than that of an isothermal halo can explain the observed N-S asymmetry in the H {\footnotesize I} thickness map in the Galaxy.

The paper is organized in the following order. \S 2 describes the formulation of the problem and models of the dark matter halo. In \S 3 we present the definition of the thickness of the H {\footnotesize I} gas and asymmetry measurements. Method and input parameters are discussed in \S 4, while \S 5 describes the results and various possible models of dark matter halo. Comparison of different models are done in \S 6. \S 7 describes a comparison with previous works. Discussion and conclusions are in \S 8 and 9 respectively.  

\section{Vertical dynamics in a Lopsided dark matter halo}

We studied the dynamics of an H {\footnotesize I} gas disk under vertical hydrostatic equilibrium in a generalized non-axisymmetric dark matter halo potential. The H {\footnotesize I} disk is gravitationally coupled to the stellar counterpart. Basically, we are going to study the equilibrium vertical structure of a two-component star-gas system in an asymmetric potential. We used the cylindrical polar coordinate system ($R$,$\varphi$,$z$) suitable for the disk geometry. Presuming that the H {\footnotesize I} gas is in hydrostatic equilibrium, the vertical dynamics of each component under the force field due to the surrounding dark matter halo can be described by coupling the Poisson equation and the equation of hydrostatic equilibrium along the normal to the mid-plane for each component. 

The Poisson equation for the system can be written as:

\begin{equation}
\frac{1}{R}\frac{\partial}{\partial R}(R\frac{\partial {\Phi_{t}}}{\partial R})  + \frac{1}{R^2} \frac{\partial^2 \Phi_{t}}{\partial {\varphi^2}} + \frac{\partial^2 \Phi_{t}}{\partial z^2}\:=\: 4\pi G \left( \sum_{i}^{2} \rho_{i}  + \rho_{h} \right) ,
\end{equation}

\noindent where $\Phi_{t}$ is the total potential due to the stars, H {\footnotesize I} gas and dark matter halo, $\rho_{i}$ with i=1 to 2 denotes the mass density for the stellar and H {\footnotesize I} components, $\rho_{h}$ is the density of dark matter halo. 


\noindent Next, we made a comparative estimate of the Poisson equation's radial and azimuthal terms (which we denote as $T_R$ and $T_{\varphi}$ respectively).
 \noindent We assumed that the potential of the Galaxy is asymmetric by a small quantity $\epsilon_{p}$ and the dominant asymmetry is in the form of a lopsidedness  (i.e.~the iso-potential contours are distributed according to $\cos{\varphi}$, corresponding to $m=1$ azimuthal wavenumber). We then wrote the total potential, $\Phi_{t}$, in the simplistic form (Rix \& Zaritsky 1995; Jog 1997):

\begin{equation}
 \Phi_{t}(R,\varphi,z)=\Phi_{0}(R,z)(1 + \epsilon_{p}\cos(\varphi -\varphi_{p})),
\end{equation}

\noindent where $\Phi_{0}(R,z)$ is the axisymmetric part of the total potential and $\varphi_{p}$ is the constant phase factor. Then the azimuthal term can be written as
\begin{equation}
T_{\varphi}= \frac{\Phi_{0}(R,z)}{R^{2}} \epsilon_{p}\cos(\varphi - \varphi_{p}),
\end{equation} 

\noindent and the radial term as: 

\begin{equation}
T_{R} \equiv  \frac{1}{R}\frac{\partial V_{c}^2}{\partial R}(1+ \epsilon_{p}\cos(\varphi - \varphi_{p})),
\end{equation}

\noindent where the rotation velocity, $V_c =\sqrt{R \partial{\Phi_{0}(R,z)}/{\partial R}}$ is determined at the mid-plane ($z=0$). To proceed further with the comparison, we need  the form of the combined axisymmetric potential $\Phi_{0}(R,z)$. We assumed that the rotation curve in the outer region of the stellar disk is mainly dominated by the dark matter halo. 

Let  $V_c \sim C_{0}R^{-\alpha}$, where the index $\alpha$ determines the shape of the rotation curve and C$_0$ is some constant of proportionality.  Note that $\alpha < 0$ i.e. a negative $\alpha$ would produce rising rotation curves usually found in dwarf galaxies. In normal spiral galaxies $\alpha \ge 0$. For example, $\alpha =0$ produces a flat rotation curve that corresponds to a screened isothermal dark matter halo ($p=1$ in our notation, see eq.[12] in \S 2.2). In the other extreme, when  $\alpha = 0.5$ the rotation curve falls in a Keplerian fashion. The dark matter halos that produce an asymptotically Keplerian rotation curve are with indices $ 1.5 < p \le 2$ (For the present work, we restrict ourselves to $p=2$). Therefore, we can say that the range of the indices in the rotation curves, $0 \le \alpha \le 1$, roughly maps to the range $1 \le p \le 2$ in the density distribution of the dark matter halos. Now clearly, for an asymptotically flat rotation curve the radial term is exactly equal to zero at the disk mid-plane ($z=0$). Whereas, the azimuthal term is given by $T_{\varphi} \sim \log(R)/R^2 \epsilon_{p} \cos(\varphi-\varphi_{p})$. For a very slowly falling rotation curve, the azimuthal term dominates over the radial term and with little algebraic manipulations it can be shown:

\begin{equation}
T_{R} + T_{\varphi} \sim [ 1 + \frac{\epsilon_{p}}{\alpha^2} \cos(\varphi-\varphi_{p})]T_{R}.
\end{equation}

\noindent Then, the Poisson equation for a disk embedded in a non-axisymmetric dark matter halo with a slowly falling rotation curve ($\alpha \neq 0$ and $\alpha \ll 1$) can be written as:
\begin{equation}
\frac{\partial^2 \Phi_{t}}{\partial z^2} \simeq 4\pi G \left( \sum_{i}^{2} \rho_{i}  + \rho_{h} \right) - [ 1 + \frac{\epsilon_{p}}{\alpha^2} \cos(\varphi-\varphi_{p})]T_{R}.
\end{equation}

\noindent Since $T_{R}$ is a negative quantity, the combined effect of the radial and azimuthal terms in the Poisson equation (Eq.[6]) is either to increase or decrease the vertical oscillation frequency of the disk depending upon the orientation of the dark matter halo and the values of $\alpha$ and $\epsilon_{p}$. In other words, this term may either try to confine the gas more towards the disk mid-plane or help flaring. 
However, the contribution from the combined radial and azimuthal terms to the thickness of H {\footnotesize I} in the outer region is not significant compared to that due to the first term in eq.[6]. The second term on the r.h.s. of eq.[6] contributes to $\sim 10\%$ in the H {\footnotesize I} thickness in the outer region for a very slowly falling rotaion curve. So in the zeroth order approximation, we can safely say that the thickness of the disk components is largely determined by the vertical pull due to the first term on the r.h.s. of the Poisson eq.[6].

\noindent The vertical hydrostatic equilibrium equation under the total potential for each component can be written as (Rohlfs 1977; Binney \& Tremaine 1987):

\begin{equation}
\sigma_{zi}^2 \frac {\partial  {\ln \rho_{i}}}{\partial z} \:=\: -\frac {\partial \Phi_{t}}{\partial z}
\end{equation}

\noindent In the above equation $\sigma_{zi}$ denotes the vertical velocity dispersion of the $i^{th}$ disk component in the problem. 
\noindent On combining the above equations (6, 7), the vertical equilibrium of each component in the disk under the dark matter halo potential is:

\begin{equation}
\frac{\partial}{\partial z}\left(\sigma_{zi}^2 \frac {\partial  {\ln \rho_{i}}}{\partial z} \right) \:=\: - 4\pi G \left( \sum_{i}^{2} \rho_{i}  + \rho_{h} \right)  + [ 1 + \frac{\epsilon_{p}}{\alpha^2} \cos(\varphi-\varphi_{p})]T_{R}.
\end{equation}


\subsection{ An analytic approximation to the thickness}

An analytic approximation for the thickness can be obtained near the disk mid-plane by direct integration of the vertical equilibrium equation. This is a useful guide to the numerical integration of eq.[8]. 

\noindent On integrating eq.[7] of the vertical equilibrium equation and using the boundary condition at z=0
$$  \rho_{i}(R,\varphi,z) = \rho_{mid}^{i}(R,\varphi,0)$$ ,
\noindent where $\rho_{mid}^{i}(R,\varphi,0)$ is the mid-plane volume density of each component, we get:

\begin{equation}
 \sigma_{zi}^2 \ln{ \left[\frac{\rho_{i}(R,\varphi,z)}{\rho_{mid}^{i}(R,\varphi,0)}  \right]}+ \Phi_{t}(R,\varphi,z) - \Phi_{t}(R,\varphi,0) \:=\: 0.
\end{equation}

\noindent Near the disk mid-plane the potential along the vertical direction will not be very different than that of the disk mid-plane. Using Taylor's expansion along the vertical direction  and the vertical equilibrium of each component, it can be shown that:

\begin{equation}
 \rho_{i}(R,\varphi,z) \:=\: \rho_{mid}^{i}(R,\varphi,0)\times e^{-\frac{z^{2}}{2 H_{i}^2(R,\varphi)}}. 
\end{equation}

\noindent Hence, the density distribution under vertical equilibrium follows a gaussian near the disk mid-plane with non-axisymmetric thickness. Here $ H_{i}(R,\varphi) \:=\: \sigma_{zi}/\nu(R,\varphi) $ and the vertical frequency $\nu$ is given by

\begin{equation}
 \nu(R,\varphi) \:=\: \sqrt{ 4 \pi G \left( \sum_{i}^{2} \rho_{mid}^{i}(R,\varphi)  + \rho_{h}(R,\varphi,0) \right)}.
\end{equation}

\noindent In deriving the above form, we  used eq.[6] with $T_{R} = 0$, since the thickness of the gas is largely determined by the vertical pull of the matter towards the disk mid-plane. The reader should bear in mind that the above analytical formula for the thickness (H$_i$) is not valid for high z.

\subsection{Models of Non-axisymmetric Dark Matter Halo}

We consider here for the first time a seven-parameter dark matter halo model for describing the neutral hydrogen gas thickness in the Milky Way. The density profile of the non-axisymmetric dark matter halo is given by: 

\begin{equation}
\rho_{h}(R,\varphi,z) \:=\: \frac{\rho_{0}(q)}{\left[ 1 + \frac{\Lambda_{\varphi}^2}{R_{c}^2(q)}\right]^p},
\end{equation}

\noindent where $\rho_{0}$ is the central mass density of the halo, $R_{c}$ is the core radius, $q$ determines the oblateness of the halo and $p$ is the index determining the nature of the density profile. The index $p=1$ denotes a softened isothermal dark matter halo with density falling as $R^{-2}$ at large radii. The name 'softened isothermal halo' is derived from the singular isothermal halo by adding a suitable core radius; these softened halos are also called screened or pseudo-isothermal halos. Since the mass $M(R)$ within a radius $R$ is proportional to $R$, the $p=1$ halo produces an asymptotically flat rotation curve. For $p=1.5$, the density $\rho_h \propto R^{-3}$ at large radii resembling the NFW halo profile (Navarro et al. 1996). The mass $M(R)$ within a radius $R$ of such a $p=1.5$ halo is proportional to $\log R$  and goes to infinity as $R$ goes to infinity but much more gradually than the $p=1$ halo.
Whereas $p=2$ denotes a perfect ellipsoid dark matter halo with density falling like $R^{-4}$ at large radii leading to essentially a finite mass halo. So the asymptotic rotation curve would be like a Keplerian one. The mass profile of the $p=2$ axisymmetric dark matter halo can be written as:

\begin{equation}
M_{2}(R) \sim 2\pi \rho_{0} R_{c}^3 q \left[\tan^{-1}(R/R_{c}) - \frac{R/R_c}{1 + R^2/R_{c}^2}\right].
\end{equation}

\noindent In the above eq.[12]

\begin{equation}
 \Lambda_{\varphi}^2 = (R_{\varphi}^2 + z^2/q^2);  R_{\varphi} = R (1 - \epsilon_{h}^{1}\cos{(\varphi - \varphi_{h})} + \epsilon_{h}^{2}\cos{2(\varphi - \varphi_{h})}),
\end{equation}

\noindent where $\Lambda_{\varphi}$ represents the surface of the concentric ellipsoid with lopsided ($m=1$) distribution superposed with second harmonics ($m=2$). The degree of lopsidedness in the dark matter distribution  is $\epsilon_{h}^{1}$  and the degree of second harmonic is determined by $\epsilon_{h}^{2}$. $\varphi_{h}$ denotes the phase of the asymmetric halo with respect to the Galactic axes. It is very hard to say without a detailed stability analysis which harmonics ($m=1$ or $m=2$) will dominate in the dark matter halo.

Since the higher harmonics ($m > 2$ or $3$) perturbations will be associated with smaller spatial scales compared to the first or second harmonics, in a collisionless dark matter halo such small scale higher harmonics are most likely to be Landau damped and large scale perturbations would be weakly damped (Weinberg 1994). Our expectation is that the dark matter halos are likely to be dominated by the first two harmonics: a lopsided ($m=1$) perturbation and an elliptical ($m=2$) perturbation. The effects of a lopsided halo perturbation and $m=2$ component in the halo perturbation onto the disk dynamics was studied by Jog (1999, 2000). Note that with $\epsilon_{h}^{2}$=0, we end up with a purely lopsided dark matter halo. With all $\epsilon_{h}=0$, the dark matter halo becomes the usual four-parameter axisymmetric halo used in previous studies (Narayan et. al. 2005; de Zeeuw \& Pfenniger 1988; Becquaert \& Combes 1997). The parameter $\varphi_{h}$ contains  important information as to how the dark matter halo is oriented with respect to the Galactic axes. It is important to note that the first two harmonics in the dark matter halo are {\it out of phase}. Writing the density distribution of the halo in the above form results in freedom to investigate a wide variety of dark matter halo potentials, while simpler to use than the triaxial ones.

\section{Thickness of the neutral hydrogen gas}
By solving the coupled Poisson equation and the vertical hydrostatic equilibrium equation (eq.[8]), we obtain the volume density of the atomic hydrogen gas $\rho_{2}(R,\varphi,z)$. To solve eq.[8] we used the mid-plane volume density and its dispersion of each component as inputs (see \S 4). We considered an exponential surface density distribution for the stellar disk with a mild lopsidedness into it.

\begin{equation}
\Sigma(R, \varphi) = \Sigma_{0} e^{-R/R_d}(1 + \epsilon_{s} \cos(\varphi - \varphi_{h})),
\end{equation}

\noindent where $\Sigma_{0}$ and R$_d$ are the central surface density and scale length of the stellar mass distribution, respectively. The value of $\epsilon_{s}$, in principle, can be determined self-consistently by calculating the response of the stellar disk to the non-axisymmetric dark matter halo considered in \S 2.2 but the task is beyond the scope of this paper and not necessary for obtaining a correct zeroth order model. Instead, we use reasonable numerical values of $\epsilon_{s}$ and calculate its effect on the H {\footnotesize I} thickness in the region of our interest ($R \ge 16$ kpc). We have checked that the gas thickness remains almost unchanged as the value of $\epsilon_{s}$ is reduced from $0.15$ to $0.0$ (corresponding to an axisymmetric stellar disk). We have also checked if $\epsilon_{s}$ alone can account for the observed systematic N-S asymmetry (prominent beyond $\sim 16$ kpc) in the H {\footnotesize I} thickness map. We found that even unrealistically high values of $\epsilon_{s}$ alone can not account for such high asymmetry in the H {\footnotesize I} flaring. Note that the contribution of the axisymmetric stellar disk itself almost drops to zero beyond $\sim 16$ kpc compared to that due to the dark matter halo. So in the present calculation, we considered only the axisymmetric stellar disk in deriving the H {\footnotesize I} thickness beyond $\sim 16$ kpc.
One important assumption in solving eq.[8] is that the velocity dispersion remains constant (isothermal approximation) along the vertical direction. 

The thickness $d(R,\varphi)$ of the gas is determined by using the second moment of the volume density distribution and is given by the following relation:

\begin{equation}
d^{2}(R,\varphi) = \frac{\int_{-\infty}^{\infty}{z^{2}\rho_{2}(R,\varphi,z) dz}}{\int_{-\infty}^{\infty}{\rho_{2}(R,\varphi,z) dz}}.
\end{equation}

\noindent $d(R,\varphi)$ gives the radial variation of the thickness along a particular azimuthal direction ({\bf $\varphi$}) in the disk. Using this thickness map, we can examine the degree of asymmetry in the gas thickness distribution.

We note that the thickness of the gas layer in this case is different from the method used by Levine et al. (2006a) but gives qualitatively similar result. Kalberla et al. (2007) calculate the gas scale height yet another way, and get result which differ both from the second moment of the distribution (eq.[16]) and from Levine et al. (2006a). All this says is that because of the exclusion of certain areas (especially $90\,^{\circ} \le \varphi \le 110\,^{\circ} $), and the possible effects of optical depth, different methods of calculating the scale height can give rise to different numerical values. In fact, it was demonstrated by Bahcall (1984) that the scale-height of a gaussian stellar distribution is roughly twice that of an expoential distribution. The important point, is that all methods show values of scale-height that are $\sim 2 - 2.5$ times larger in the northern part of the Galaxy ($l=0 - 180\,^{\circ}$) than the south ($l=180 - 360\,^{\circ}$) and it is this difference that we trying to model and explain. 


\noindent Let $A(\varphi_{k})$ be the area under the thickness curve along a particular azimuthal angle ($\varphi_{k}$):

\begin{equation}
A(\varphi_{k}) = \int_{R_{min}}^{R_{max}}{d(R,\varphi_{k}) dR}
\end{equation}

\noindent In the above equation $R_{min}$ and  $R_{max}$ represent the initial and the final radius respectively in the available observation of the gas thickness. The degree of asymmetry in the thickness distribution, presuming that we are considering a smooth curve, can then be described by:

\begin{equation}
\eta(\varphi_{k}) = \frac{\left|{A(\varphi_{k}) -  A(\varphi_{k} +\pi)}\right|}{ A(\varphi_{k}) + A(\varphi_{k} +\pi)}.
\end{equation}

\noindent The range of $\eta$ is $0 \le \eta \le 1$, with $\eta$=0 denoting a symmetric distribution. On the other hand $\eta$=1 denotes a highly asymmetric gas/star distribution leading to a one-sided flaring in the galaxy. It is unlikely that $\eta \approx 1$, because this would mean that the velocity dispersion of the gas in one half of the galaxy is almost zero (cold component) compared to the other half. This may lead to an instability in the disc. Considering $\varphi_{k} = 90\,^{\circ}$, the above relation would produce the numerical value of the observed N-S asymmetry in the thickness map.
   
\section{Method and Input parameters}
The present study focuses on deriving the thickness map of the neutral hydrogen gas (H {\footnotesize I}) in the  very outer region ($R \ge 16$ kpc) of the Galactic disk. Since of the various baryonic components the stars and  H {\footnotesize I} dominate the mass in this region, we neglect the effect of molecular hydrogen gas (H$_2$) on the thickness distribution of H {\footnotesize I}. Eq.(8), which describes the zeroth order vertical equilibrium under a non-axisymmetric dark matter distribution, represents two coupled differential equations for the two disk components: H {\footnotesize I} and stars. The vertical density distribution for each component, responding to the total potential due to the disk and the dark matter halo, was solved numerically as an initial value problem using the fourth order Runge-Kutta method of integration (Press et al. 1994). The details of this method are presented in Narayan \& Jog (2002) and Narayan et al. (2005). Because of the underlying non-axisymmetry, eq.(8) is solved for each azimuthal direction along the Galactocentric radius. Along a particular azimuth, at a particular radius, we calculate the second moment of the vertical density distribution for each component according to eq.(16) and call it the thickness of that component. Repeating this procedure for regular intervals in the azimuth along the galactocentric radius give us the thickness map of the neutral hydrogen gas layer in the Galaxy.

The primary input parameters needed for the method are the mid-plane volume density and the vertical velocity dispersion for each disk component. The H {\footnotesize I} mid-plane volume density is obtained from the LAB survey (Kalberla et al. 2005; Levine et al. 2006a). For the stellar disk, we used its surface density according to eq.(15) for which we need to know the central surface density and scale length. Note that eq.[15] represents a lopsided stellar disk and as discussed in \S 3, the value of $\epsilon_s$ does not effect much the H {\footnotesize I} thickness in the very outer region of the Galaxy. So we considered effectively an axisymmetric stellar disk in the region of our interest. Using the following measured/inferred quantities: the stellar surface density at the solar region $\Sigma_{\odot}$, the disk scale length $R_d$ and the distance of sun from the Galactic center R$_{\odot}$, we can derive the central surface density $\Sigma_{0}$ and infact the surface density at any radius. We use $\Sigma_{\odot}$ = 45 M$_{\odot} {\rm pc}^{-2}$ which is consistent with 48 $\pm$ 9 M$_{\odot} {\rm pc}^{-2}$ obtained by Kuijken \& Gilmore (1991) and 52 $\pm$ 13 M$_{\odot} {\rm pc}^{-2}$ obtained by Flynn \& Fuchs (1994) for the total surface density, after the gas density is subtracted. We use the IAU recommended value for $R_{\odot}$ (=8.5 kpc) and this was also  used to determine the observed thickness map for H {\footnotesize I} gas (Levine et al. 2006a). The scale length $R_d$ was set equal to 3.2 kpc (Mera et al. 1998) in accordance with the recent determinations of smaller disk scale-length for our Galaxy.

\subsection{Stellar and H {\footnotesize I} velocity dispersion}

The stellar vertical dispersion was derived from observation of radial dispersion by Lewis \& Freeman (1989) and then using the assumption that the ratio of the vertical to radial velocity dispersion is equal to 0.45 at all radii in the Galaxy, equal to its observed value in the solar neighbourhood as obtained from the analysis of the Hipparcos data (Dehnen \& Binney 1998, Mignard 2000).

The H {\footnotesize I} velocity dispersion ($\sigma_{H {\footnotesize I}}$) has been observed to be nearly constant with radius at approximately 9$\pm$1 km s$^{-1}$ (Spitzer 1978; Malhotra 1995) in the inner Galaxy (out to the solar circle). Beyond the solar circle, however, the dispersion has not yet been measured.  A study of 200 external galaxies (Lewis 1984) shows that the observed dispersion has a very narrow range, about 8$\pm$1 km s$^{-1}$, consistent with observations of our Galaxy. Sicking (1997) showed that in two external galaxies, dispersion decreases slowly out to the outer edge of the H {\footnotesize I} layer. In a number of other galaxies, the velocity dispersion decreases and then stabilizes at a constant value of 7$\pm$1 km s$^{-1}$ (Shostak \& van der Kruit 1984; Dickey 1996; Kamphuis 1993). This decrease in velocity dispersion is perhaps due to the lesser number density of supernovae (SN) in the outer region (McKee \& Ostriker 1977). On the other hand, recent work by Dib et al. (2007) shows that the SN do not affect the observed gas velocity dispersion in the galactic outskirt. A large fraction of the observed velocity dispersion is non-thermal (or turbulent) in origin and the supernovae could be the major source for this turbulent nature of the H {\footnotesize I} velocity dispersion. Note that the thermal contribution accounts for only about 1 km s$^{-1}$ (Spitzer 1978). In any case, the H {\footnotesize I} velocity dispersion is as crucial as the dark matter distribution is in determining the H {\footnotesize I} thickness map in the outer Galaxy. Unfortunately, in the absence of any direct measurement of the H {\footnotesize I} velocity dispersion ($\sigma_{HI}$) beyond the solar circle, we use $\sigma_{HI}$ as a model parameter in the fitting problem. We construct various model based on the H {\footnotesize I} velocity dispersion values and models of dark matter distribution (Eq.[12]).

\subsection{Model rotation curves}

We model the rotation curve for the Galaxy using a bulge, an exponential disk for the stars and gas and a dark matter halo. Modelling the observed rotation curve is a non-trivial task given that there are various kind of uncertainties and turns out to be a non-linear regression problem in a multidimensional parameter space. Naturally, such modelling would suffer from uniqueness problem. So we aim here to reproduce the main features in the observed rotation curve and try to generate such rotation curves for which the circular speed lies in the range determined by the relation $\Theta_{\circ}=(27 \pm 2.5) R_{\circ}$ km$s^{-1}$ due to Kerr \& Lynden-Bell (1986). Since we use the IAU recommended value for $R_{\circ}=8.5$ kpc, the above range implies $V_c = 230 \pm 21$ km$s^{-1}$ at the solar radius. Keeping these constraints in mind, we proceed to derive the model rotation curves in the following way. We assume that the disk and the bulge are aligned with the symmetry axis of the dark matter halo. We also assume the virial equilibrium in the Galaxy. Then the square of the total circular velocity in the disk mid-plane ($z=0$) can be written as: 

\begin{equation}
V_{c}^2 = V_{bulge}^2 + V_{stars}^2 + V_{gas}^2 + V_{dmh}^2,
\end{equation}    

where we adopt a Plummer-Kuzmin bulge model to derive the bulge contribution to the rotation curve. The density profile of the bulge is given by the following formula (Binney \& Tremaine 1987):

\begin{equation}
\rho_{b}(R)=\frac{3M_{b}}{4\pi R_{b}^{3}}\left(1+\frac{R^{2}}{R_{b}^{2}}\right)^{-5/2},
\end{equation}

\noindent where $R_{b}$ is the bulge scale-length and $M_{b}$ is the total bulge mass. We have used same values of these two parameters in all of our model rotation curves e.g. $R_{b}=2.5 $ kpc and $M_{b}=2.8 \times 10^{10} M_{\odot}$ (see Blum 1995). The inclusion of this simple spherical bulge reproduces a reasonable looking rotation curve for the Galaxy. However, the contribution of the bulge to the H {\footnotesize I} thickness in the region of interest ($R \ge 16$ kpc) is almost negligible, the bulge contributes to $\sim 5 \%$ to the overall gas thickness.

The disk contribution to the circular speed is derived using the potential for the exponential mass distribution ($\Sigma(R)=\Sigma_{0}e^{-R/R_d}$ where values of the parameters are mentioned in \S 4) and similarly for the various dark matter halo models for which the parameters are mentioned in the appropriate places below. The gas contribution to the total circular speed is derived from the following exponential distribution for the H {\footnotesize I} beyond $14$ kpc (Levine et al. 2006a):

$$ \Sigma_{HI}(R)=4.5\times e^{[-(R - 14 kpc)/4.3 kpc]} $$  

Below 14 kpc, we use a constant surface density for the H {\footnotesize I} gas (as observed) to derive its contribution to the rotation curve. Although there is a notable difference of our model rotation curves with the observed one (Brand \& Blitz 1993), the slowly falling rotation curves due to the $p=2$ halo model follow the trend found in the Milky Way's rotation curve in the very recent analysis of Xue et al. (2008) based on the SDSS data.  

\section{H {\footnotesize I} flaring and nature of dark matter halo}

There is a clear North-South asymmetry in the thickness map of the H {\footnotesize I} gas in our Galaxy. It is not obvious what would have caused such asymmetry in the H {\footnotesize I} gas distribution. What is the underlying nature of this asymmetry? It remains to be shown whether this is purely a gas dynamical effect or reflects some gravitational effect. Below we describe a step by step analysis of the cause of this asymmetry, which gradually reveals the nature of dark matter halo in our Galaxy. We used averaged flaring data over the north ($0\,^{\circ}$ $\le \phi \le$ $180\,^{\circ}$) and south ($180\,^{\circ}$ $\le \phi \le$ $360\,^{\circ}$) respectively excluding about $15\,^{\circ}$ region about the Sun-Galactic centre line. The vertical volume density distribution of the gas is derived using the rotation curve due to Brand \& Blitz (1993). The thickness of the gas is then derived by taking the second moment of the density distribution after the thickness filter has been applied. In this respect, the thickness measurement of the gas is different from the Levine et al. (2006a) who uses the tail integration method and it is also different from Kalberla et al. (2007) who uses half width at half maximum (HWHM) for the half-thickness of the gas later. In all our analyses below, we use the same thickness map for the gas, although it is understood that a different rotation curve would indeed produce different thickness map for the gas. However, on doing an error analysis we find that the error in calculating the distance due to an error in the rotation velocity is small ($\sim 10\%$ at large galactocentric distances) to produce an appreciable change in the observed thickness of the gas. The readers are referred to \S 8 for a discussion on the dependence of the thickness data on the rotation curve. The observed N-S asymmetry denotes the average asymmetry in the thickness map of the H {\footnotesize I} gas and according to eq.[18], its given by $\eta_{obs}=0.262$.  

\subsection{Models of axisymmetric dark matter halo}
 
We first considered a simple axisymmetric dark matter halo model (with all $\epsilon_{h} = 0$ in eq.[14]) for the Galaxy and used different velocity dispersions for the H {\footnotesize I} in the two halves to see if the observed N-S asymmetry could be reproduced.

 \subsubsection{\bf{$p=1$, Softened Isothermal Halo}}
The softened isothermal dark matter halo produces naturally the asymptotically flat rotation curve in a spiral galaxy. It has been shown previously based on the Wouterloot et al. (1990) data that a $p=1$ softened isothermal halo of any shape (oblate or prolate) cannot explain the observation (Narayan et al. 2005). In the present study, we found the same trend, so we considered a nearly spherical halo to begin with for further investigation. We adopt the parameters of the $p=1$ isothermal dark matter halo from the mass model of our Galaxy based on microlensing observation by Mera et al. (1998). We found that the $p=1$ halo with core radius $R_c$= 5 kpc, central density $\rho_{0}$=0.035 M$_{\odot}$pc$^{-3}$, which gives rise to an asymptotically flat rotation curve with terminal velocity of 220 km s$^{-1}$ (Brand \& Blitz 1993) (see bottom right of Fig.~1), and with $\sigma_{H {\footnotesize I}}$=9 km s$^{-1}$ cannot explain the present observation (LAB data). The fact that $p=1$ softened isothermal halo can't explain the flaring in the gas thickness distribution has already been verified by Narayan et al. (2005) in the case of Wouterloot et al. (1990) data and more recently by Kalberla et al. (2007) based on the LAB survey data. Now, there are two aspects of the H {\footnotesize I} thickness map derived from the LAB survey data: the radial variation of the gas thickness in the North and the South and the prominent asymmetry between the North and the South as mentioned already. An axisymmetric $p=1$ halo with the above mentioned parameters (giving rise to $V_c = 220$ km s$^{-1}$) and $\sigma_{H {\footnotesize I}} = 9$ km s$^{-1}$ appears to be quite massive and hence difficult to reproduce the present observation. By inspection of Fig.~1 (top left), it is easy to check that there has to be an order of magnitude decrease in the dark matter mid-plane density to reproduce the observed gas thickness at 30 kpc. Moreover, the striking difference between the two slopes makes it clear that even a linearly increasing velocity dispersion with radial distance (although unphysical) cannot fit the observation. Obviously, an axisymmetric $p=1$ halo alone can not explain the other important aspect of the thickness map, namely, the observed N-S asymmetry. This fact leads us to explore another possibility of a model of an axisymmetric $p=1$ halo accompanied by a non-axisymmetric distribution of H {\footnotesize I} velocity dispersion. It can be explained in a simple way that a different H {\footnotesize I} velocity dispersion on both sides of the hemispheres is not going to improve the situation either. In a naive theory, the thickness of H {\footnotesize I} gas can be written as $h_{H {\footnotesize I}} \propto \sigma_{H {\footnotesize I}}/\sqrt{G \rho_{mid}}$. Using this simple formula, it is easy to see that in order to increase the thickness by a factor of 2 at a particular radius, one needs to raise the dispersion by a factor of $\sim$ 2 (because $\rho_{mid} $ remains the same). At $R = 30$ kpc, the thickness of H {\footnotesize I} gas in the Northern hemisphere is roughly a factor of 2 more than that in the Southern hemisphere and to explain this asymmetry based on purely gas dynamical effect and the axisymmetric $p=1$ isothermal halo (giving rise to a flat rotation curve), one needs to have $\sigma_{H {\footnotesize I}} \sim 20$ km s$^{-1}$ which is unlikely according to the standard models of ISM of our Galaxy.

\begin{figure*}
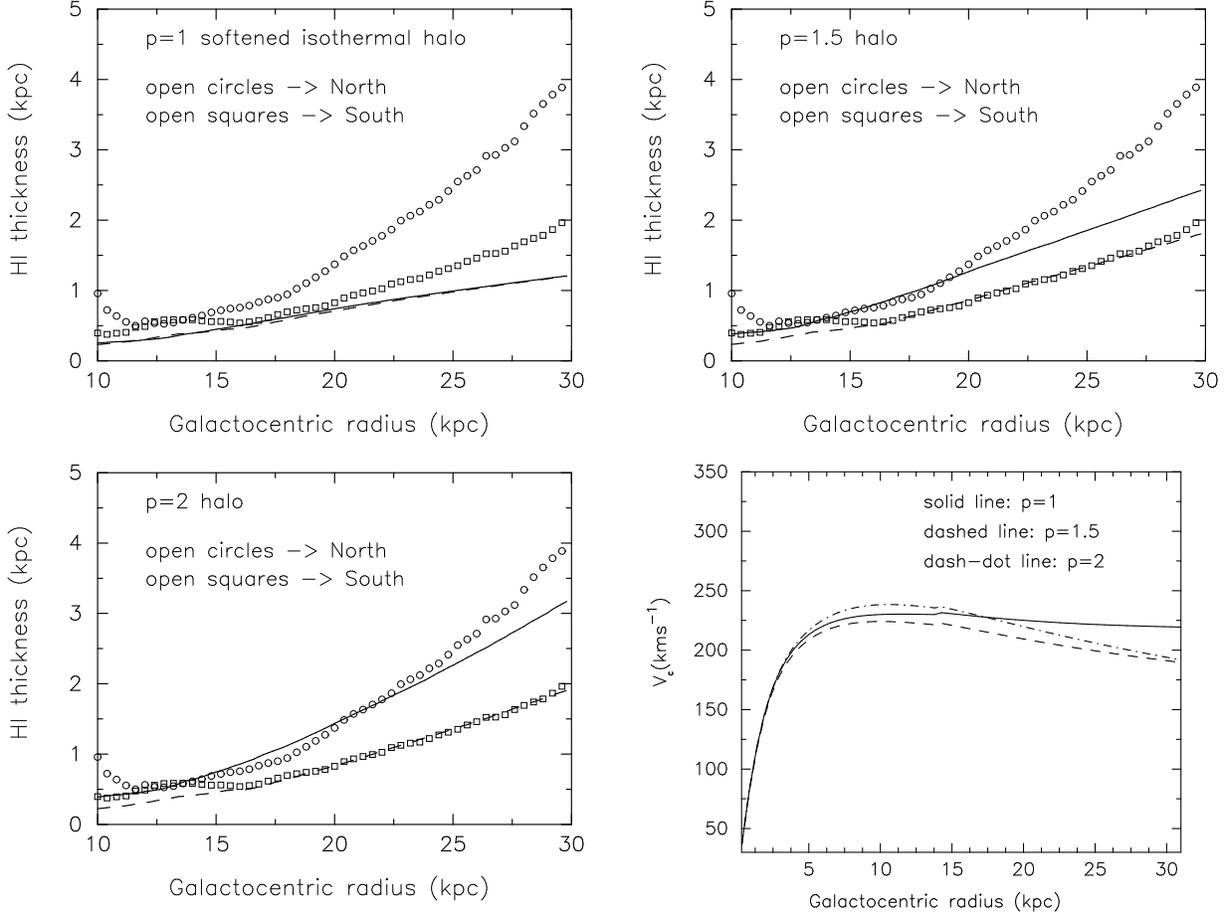
{}
  \newlength{\figwidth}
  \setlength{\figwidth}{\textwidth}
  \addtolength{\figwidth}{-\columnsep}
  \setlength{\figwidth}{0.5\figwidth}
  
  \begin{minipage}[t]{\figwidth}
    \mbox{}
    \vskip -7pt
    \centerline{\includegraphics[width=0.72\linewidth,angle=270]{fig1a.eps}}

  \end{minipage}
  \hfill
  \begin{minipage}[t]{\figwidth}
    \mbox{}
    \vskip -7pt
    \centerline{\includegraphics[width=0.72\linewidth,angle=270]{fig1b.eps}}

  \end{minipage}
\begin{minipage}[t]{\figwidth}
    \mbox{}
    \vskip -7pt
    \centerline{\includegraphics[width=0.72\linewidth,angle=270]{fig1c.eps}}

  \end{minipage}
\begin{minipage}[t]{\figwidth}
    \mbox{}
    \vskip -7pt
    \centerline{\includegraphics[width=0.74\linewidth,angle=270]{fig1d.eps}}

  \end{minipage}

    \caption{Half thickness of neutral hydrogen gas (first three panels) in the Milky Way and rotation curve models (bottom right) for different dark matter halos. Top left: Solid line is the model fitted to the North and dashed line to the South for the $p=1$ axisymmetric halo giving rise to a flat rotation velocity of 220 km s$^{-1}$ at large radii. Top right: Model half thickness due to axisymmetric $p=1.5$ dark matter halo with different $\sigma_{H {\footnotesize I}}$ in the two halves. Bottom left: Model half thickness due to axisymmetric $p=2$ dark matter halo with different $\sigma_{H {\footnotesize I}}$ in the two halves}

\end{figure*}

\subsubsection{\bf{p=1.5, an NFW type halo}}
At large distances ($R \gg R_c$) from the centre, $p=1.5$ halo resembles an NFW profile (Navarro et al. 1996) for the dark matter halo. In our model of vertical equilibrium, this seems to be preferable compared to the $p=1$ softened isothermal dark matter halo. Because of the density falling faster than the $p=1$ isothermal halo, the resulting rotation curve also starts falling beyond  about 10 kpc. We use eq.[2.96b] of Binney \& Tremaine (1987) to generate the rotation curve of the $p=1.5$ halo. The axisymmetric p=1.5 halo with $R_c$= 8 kpc and $\rho_{0}$=0.025 M$_{\odot}$pc$^{-3}$ produces a reasonable rotation curve (see bottom right of Fig.~1). At $R_\odot$ the rotation velocity is 223 km s$^{-1}$ and at $2R_\odot$, $V_c$=216 km s$^{-1}$. At 25 kpc, the difference in rotation velocities between the $p=1$ and $p=1.5$ halo is approximately 20 km s$^{-1}$. In the top right panel of Fig.~1, we show the half-thickness of H {\footnotesize I} gas due to the $p=1.5$ halo model considered here. The axisymmetric halo model with $\sigma_{H {\footnotesize I}}$=9.2 km s$^{-1}$ fits well the observation in the southern part beyond about 15 kpc. However, the same model does not fit the thickness curve at all in the northern halves. The solid line shows the curve with H {\footnotesize I} velocity dispersion $\sim$ 12.2 km s$^{-1}$ and yet does not give a good fit. An increase in the H {\footnotesize I} velocity dispersion beyond 12 km s$^{-1}$ would force the model curve to intersect the observed one only at a single point in the North. This again demonstrates that the N-S asymmetry in the thickness map of the H {\footnotesize I} is probably not due to a gas dynamical effect, rather it arises due to some gravitational effect.

\subsubsection{\bf{$p=2$, a Perfect Ellipsoidal Halo}} 
Previous studies by Narayan et al. (2005) have investigated the axisymmetric $p=2$ halo in considerable detail to explain the H {\footnotesize I} thickness in the Galaxy. Their $p=2$ halo provides a good fit to the Wouterloot et al. (1990)  flaring data. In the present study, we also considered an axisymmetric p=2 halo and see if different H {\footnotesize I} velocity dispersions can explain the N-S asymmetry. We consider a core radius $R_c$= 9.4 kpc and central density $\rho_{0}$=0.035 M$_{\odot}$pc$^{-3}$ which produces a reasonable rotation curve (see Fig.~1) within the uncertainties in the observation out to 30 kpc where the difference between the $p=1$ (flat rotation curve) and $p=2$ rotation curves is $\sim 25$ km s$^{-1}$. Note that the rotation curves for both $p=1.5$ and $p=2$ halos are slowly falling with radial distance. Recent analysis of the kinematics of a large number of Blue Horizontal-Branch halo stars from the SDSS database by Xue et al. (2008) show that the rotation curve of our Milky Way is actually falling slowly with the Galactocentric radius. So this direct observational analysis supports the falling trend in the rotation curve of Milky Way. Infact, the choice of the halo parameters for the p=2 case makes the rotation curve more realistic as observed. At 25 kpc, the circular speed due to our $p=2$ halo differs from the flat rotation curve only by $\sim 15$ km s$^{-1}$ which is well within the observed error bars. Beyond about 15 kpc, with linearly decreasing H {\footnotesize I} velocity dispersion from 9.2 (at 10 kpc) to 8.0 (at 30 kpc)  km s$^{-1}$, the $p=2$ halo considered here gives a good fit to the observed data in the southern halves. However, even with $\sigma_{H {\footnotesize I}}$=12.2 or 14.2 km s$^{-1}$, the $p=2$ halo does not fit well the observation in the north. With 14.2 km s$^{-1}$, the model curve over-estimates the observation between 13 - 23 kpc and underestimates beyond 24 kpc. On the other hand with 12.2 km s$^{-1}$, the model under-estimates the observation beyond 20 kpc. The solid curve in Fig.~1 (Bottom left panel) is with a constant $\sigma_{H {\footnotesize I}}$=13.2 km s$^{-1}$ in the Northern halves. Compared to $p=1$ and $p=1.5$, an axisymmetric $p=2$ halo with different H {\footnotesize I} velocity dispersion comes closer to the observation. We have also tried to use a linearly decreasing H {\footnotesize I} velocity dispersion from 13.2 or 12.2 km s$^{-1}$, but no good fit found. It becomes quite clear that the observed H {\footnotesize I} flaring in the North cannot be explained with any physically meaningful variation of the H {\footnotesize I} velocity dispersion. 

\subsection{Non-axisymmetric dark matter distribution}

Based on the above three axisymmetric cases we have investigated so far, we conclude that an asymmetric state of ISM combined with axisymmetric dark matter potential is not able to explain the observed asymmetry in the thickness map indicating that the observed asymmetry is probably not due to a gas dynamical effect. The large scale asymmetry in the north-south thickness map is instead likely to be gravitational in origin. Beyond 16 kpc, the stellar contribution is not significant compared to the dark matter, which implies that the N-S asymmetry is not likely due to the asymmetric stellar disk and also the observed asymmetry in the stellar disk is not very high in the Galaxy. This gives us a room to explore whether the N-S asymmetry originates due to an asymmetric dark matter halo. As explained in \S 1, such asymmetric dark matter halos are not uncommon even in the cosmological N-body simulations. By inspecting the observed H {\footnotesize I} thickness map, we guess that the N-S asymmetry is primarily lopsided ($m=1$) in nature. A lopsided dark matter halo has been used to provide an explanation for the observed lopsidedness in the underlying stellar disk and for the asymmetry in the rotation curves of disk galaxies (Jog 1997, 2002). Below we test our hypothesis that the Milky Way's dark matter halo is lopsided. We show next that the observed nature of the H {\footnotesize I} distribution demands that the lopsided dark matter halo to be oriented with phase angle $\phi_h=270\,^{\circ}$. In the present case, $\phi_h$ is the angle between the direction in which the dark matter distribution is elongated more on one side compared to the other and the Galactocentric coordinate axes. So the dark matter density maximum of the lopsided halo is along the southern direction ($\phi =270\,^{\circ}$) and the density minimum is along the northern direction ($\phi =90\,^{\circ}$). We test $p=1.5$ and $p=2$ halos as the likely candidates in our further investigation; $p=1$ is not used because it fails to produce the observed flaring even in the south.

\begin{figure}[h]
\includegraphics[angle=270,scale=0.60]{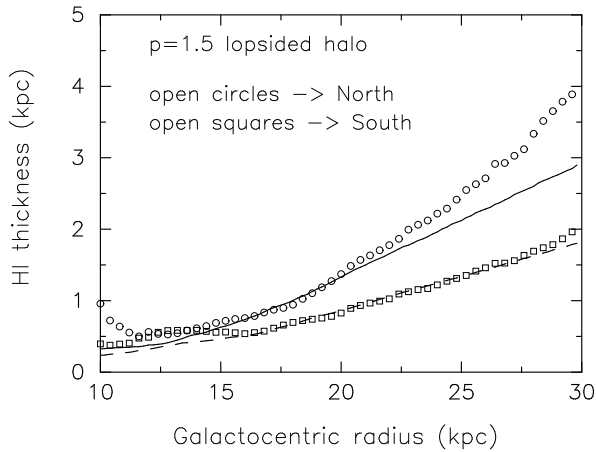}
\caption{Half thickness of neutral hydrogen gas in Milky Way. Solid line is the model fitted to the North and dashed line to the South. The value of H{\footnotesize I} velocity dispersion $\sigma_{H {\footnotesize I}}$ is 9.2 kms$^{-1}$.}
\end{figure}

\subsubsection{\bf {p=1.5 lopsided halo}}
Having oriented the dark matter halo with phase $\phi_h=270\,^{\circ}$, we found that an axisymmetric ($\epsilon_h$=0) $p=1.5$ halo with H {\footnotesize I} velocity dispersion 9.2 km s$^{-1}$ fits quite well the observation in the southern halves (see Fig.~1). However, a purely lopsided configuration ($\epsilon_{h}^{2} = 0$ in eq.[14]) with reasonable parameters can not explain the observed asymmetry in the thickness map of H {\footnotesize I} gas. The parameter ranges we have tried are $\sigma_{H {\footnotesize I}}= 7 - 10$ km s$^{-1}$ and  $\epsilon_{h}^{1} = 0.02 - 0.4$. We found that no combination of these parameters could provide a good fit to the observed data. Further, we tried to fit the combined model of the dark matter halo (given in eq.[12]) with lopsidedness as the dominant component. Again, for the above mentioned parameter range no good fit to the observed data was found. Fig.~2 shows a configuration of the dark matter halo in which both the first ($m=1$) and second ($m=2$) harmonics are equal in strength and out of phase with each other. Of course, such configuration of $p=1.5$ halo with reasonable H {\footnotesize I} velocity dispersion ($9.2$ km s$^{-1}$) provides better fit to the observation than the axisymmetric configuration of $p=1.5$ halo with high $\sigma_{H {\footnotesize I}}$.
\begin{figure*}[h!]
  \setlength{\figwidth}{\textwidth}
  \addtolength{\figwidth}{-\columnsep}
  \setlength{\figwidth}{0.5\figwidth}
  
  \begin{minipage}[t]{\figwidth}
    \mbox{}
    \vskip -7pt
    \centerline{\includegraphics[width=0.72\linewidth,angle=270]{fig3a.eps}}

  \end{minipage}
  \hfill
  \begin{minipage}[t]{\figwidth}
    \mbox{}
    \vskip -7pt
    \centerline{\includegraphics[width=0.72\linewidth,angle=0]{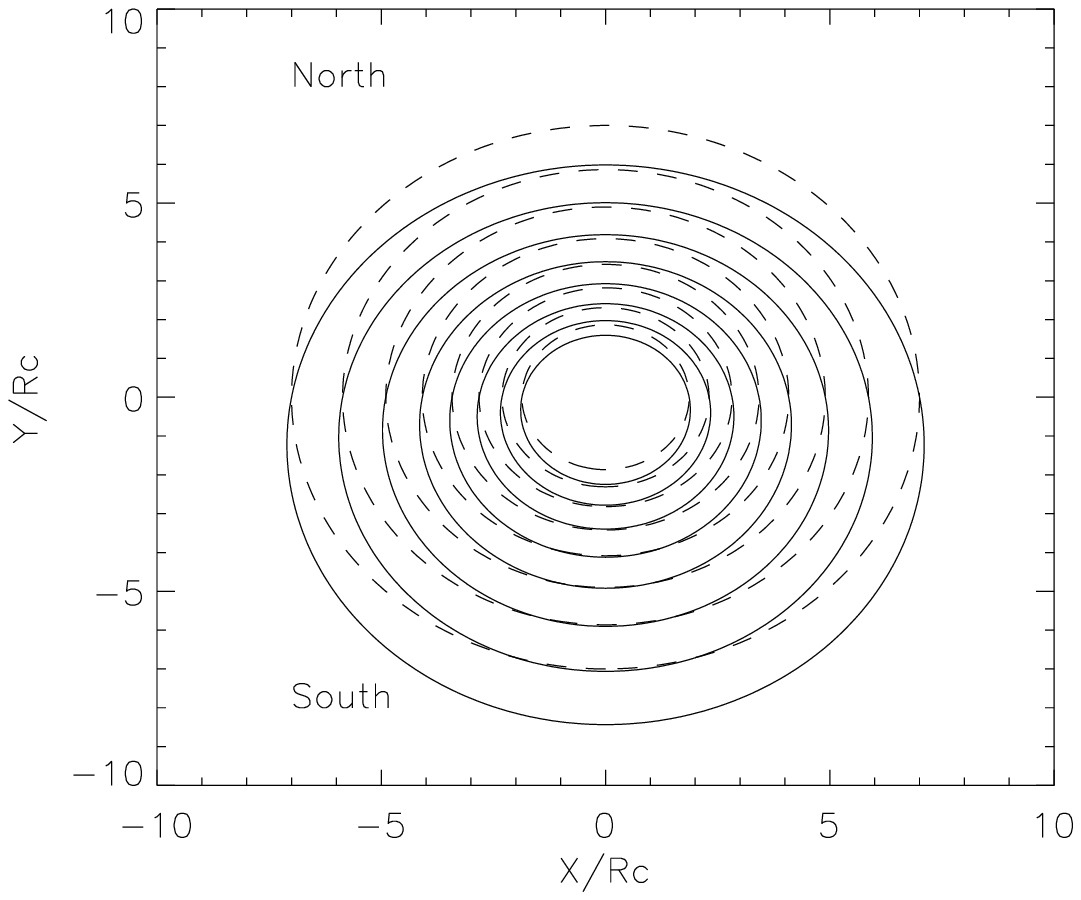}}

  \end{minipage}

  \begin{minipage}[t]{\figwidth}
    \mbox{}
    \vskip -7pt
    \centerline{\includegraphics[width=0.72\linewidth,angle=270]{fig3c.eps}}

  \end{minipage}

\caption{ Left: Half thickness of neutral hydrogen gas in Milky Way, {\bf model p2L}. Solid line is the model fitted in the North and dashed line to the South. See Table 1 for the model parameters. Right: Density contours of the $p=2$ lopsided dark matter halo. dashed lines are for the axisymmetric perfect ellipsoidal ($p=2$) halo. Solid lines are for the purely lopsided $p=2$ dark matter halo in case of {\bf model p2L}. R$_c$ is the core radius of the dark matter halo. Contour levels are 0.05$\times \rho_{0}$ for the inner most one and decreasing by a factor of 2 outwardly. H{\footnotesize I} velocity dispersion $\sigma_{H {\footnotesize I}}$ is 11.0 kms$^{-1}$ for this model.}

\end{figure*}

\subsubsection{\bf{p=2 lopsided halo}}
Since the axisymmetric $p=2$ halo already seemed promising compared to others, we carried out a detailed analysis of the $p=2$ lopsided halo here. We used the basic parameters like core radius and central density of the axisymmetric $p=2$ configuration (see \S 5.1.3). In order to find a best fit model of the Galaxy under a purely lopsided halo, we make a 2D grid of two independent and free parameters ($\sigma_{H {\footnotesize I}}$, $\epsilon_{h}^{1}$) spanning a large dynamical range i.e. [$\sigma_{H {\footnotesize I}}$, $\epsilon_{h}^{1}$]=[(7 - 12) km s$^{-1}$, (0.05 - 0.40)] for this halo model. We found that with $\sigma_{H {\footnotesize I}} \le 9.0$ km s$^{-1}$, a purely lopsided $p=2$ halo is not sufficient to explain the observation. With some more exploration, we found that a purely lopsided p=2 halo with $\epsilon_{h}^{1} =0.17$ and  $\sigma_{H {\footnotesize I}} = 11$ km s$^{-1}$  does fit the observation quite well (see Fig.~3). We call this configuration model p2L. The rotation curve for this model is shown in the bottom panel of Fig.~3. The difference between the rotation velocities in the North and the South is $\sim 20$ km s$^{-1}$ for this model. The density contours of the dark matter halo are shown in the right panel of Fig.~3. Note that the H {\footnotesize I} velocity dispersion for this model p2L is fairly high. So we further explored the parameter space consisting of $\epsilon_{h}^{1}$, $\epsilon_{h}^{2}$ and $\sigma_{H {\footnotesize I}}$ for the $p=2$ dark matter halo to find the best possible model to explain the observed radial variation of the thickness map and its N-S asymmetry. 

Based on the  different values of H {\footnotesize I} velocity dispersion and different values of the $\epsilon_h^{1}$ and $\epsilon_h^{2}$, we construct three models (model A, model B and model C) which give a very good fit to the averaged observation in both halves of the Galaxy and thereby explain the observed asymmetry.

\noindent{\bf {Model A}}

We assumed the velocity dispersion of H {\footnotesize I}, $\sigma_{H {\footnotesize I}} = 8.5$ km s$^{-1}$, to be constant out to 30 kpc for simplicity. The value of the velocity dispersion is more reasonable one and it is very closed to what has been used by Kalberla et al. (2007) in their best fit model. Once $\sigma_{H {\footnotesize I}}$ is fixed, we are essentially left with again a 2D grid of free parameters $\epsilon_{h}^{1}$ and $\epsilon_{h}^{2}$ for the dark matter halo. The parameter range explored is [$\epsilon_{h}^{1}$, $\epsilon_{h}^{2}$]=[(0.05 - 0.40), (0.05 - 0.40)]. We found that the inclusion of the second harmonic component in the dark matter halo dramatically improved the quality of the fit compared to all the previously explored cases. The best fit model parameters, for the above mentioned $\sigma_{H {\footnotesize I}}$, are $\epsilon_{h}^{1}$=0.20 and $\epsilon_{h}^{2}$=0.18. Such configuration of the dark matter halo which is indeed lopsided, reproduced the N-S asymmetry quite well in the Galaxy (see Fig.~4). The rotation curves for the model in the two halves of the Galaxy are shown in the right panel of Fig.~4. The circular speeds at the solar radius are quite comparable to the $p=1$ flat rotation curve within the uncertainties. At 25 kpc, the north and south rotation curves differ from each other by $\sim 19$ km$^{-1}$, the value in the north being $\sim 209$ km$^{-1}$. The density contours of the model dark matter halo are shown in the bottom panel of Fig.~4. Because of the presence of the second harmonics ($m=2$), the contours are elongated in the East-West direction of the Galaxy too. It is quite clear now that a simple model that consists only of a lopsided dark matter halo does not  reproduce the observed asymmetry in the thickness map of the atomic hydrogen gas in our Milky Way. The lopsided dark matter halo with some amount of second harmonic ($m=2$) superposed onto it appears to give the best fit to the data compared to the purely lopsided dark matter halo. We call the emerging picture of dark matter halo as elliptically perturbed lopsided dark matter halo.
   
\begin{figure*}
  \setlength{\figwidth}{\textwidth}
  \addtolength{\figwidth}{-\columnsep}
  \setlength{\figwidth}{0.5\figwidth}
  
  \begin{minipage}[t]{\figwidth}
    \mbox{}

    \centerline{\includegraphics[width=0.82\linewidth,angle=270]{fig4a.eps}}

  \end{minipage}
  \hfill
  \begin{minipage}[t]{\figwidth}
    \mbox{}

    \centerline{\includegraphics[width=0.82\linewidth,angle=270]{fig4b.eps}}

  \end{minipage}

\begin{minipage}[t]{\figwidth}
    \mbox{}

    \centerline{\includegraphics[width=1.1\linewidth,angle=0]{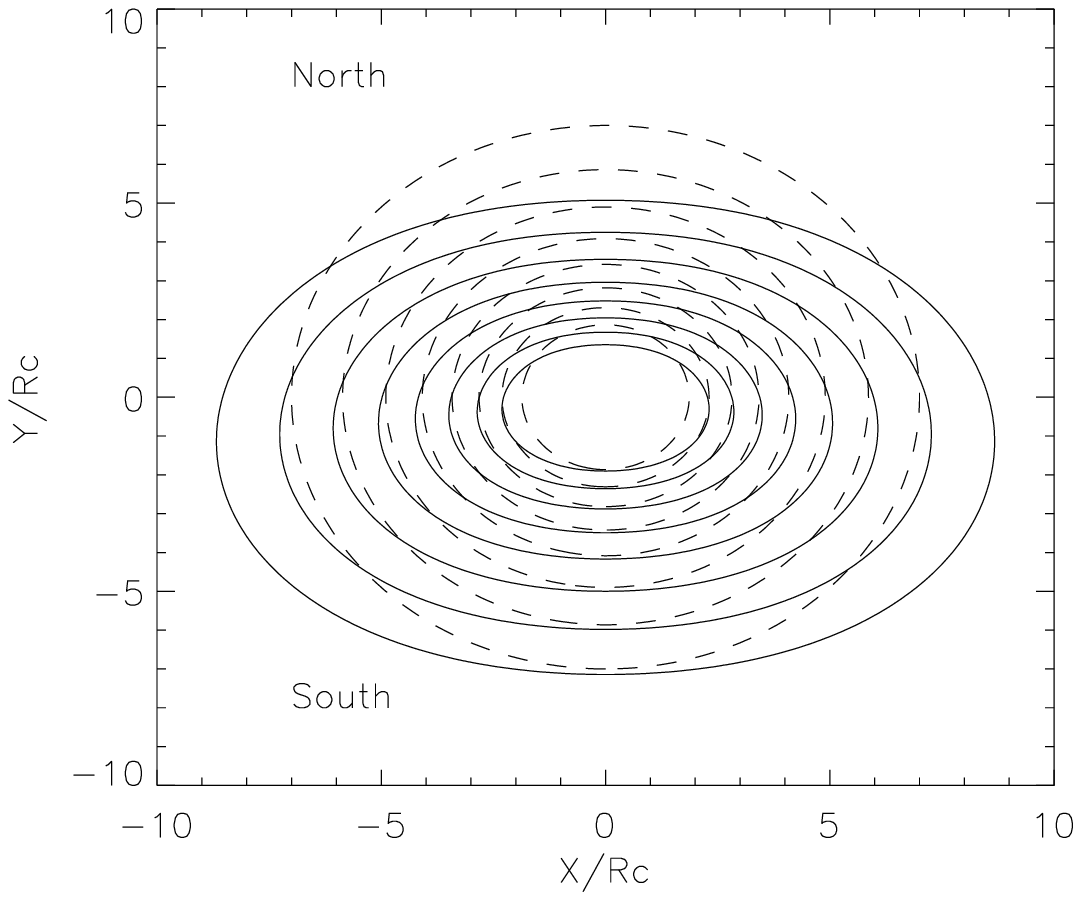}}

  \end{minipage}

    \caption{Top left: Half thickness of neutral hydrogen gas in Milky Way, {\bf model A}. Solid line is the model fitted in the North and dashed line to the South. Beyond $\sim$ 16 kpc the fit is quite good. The model parameters are given in Table 1. Top right: Rotation curve of the Galaxy for {\bf model A}. Bottom: Density contours of the dark matter halo of {\bf model A}. dashed lines are for the axisymmetric perfect ellipsoidal (p=2) halo. Solid lines are for the elliptically perturbed lopsided p=2 dark matter halo with second harmonics ($m=2$) being $\sim$ the first ($m=1$) one (see table 1). R$_c$ is the core radius of the dark matter halo. Contour levels are 0.05$\times \rho_{0}$ for the inner most one and decreasing by a factor of 2 outwardly. $\sigma_{H {\footnotesize I}}$ used for this model is 8.5 kms$^{-1}$. }

\end{figure*}
\begin{figure*}{}
  \setlength{\figwidth}{\textwidth}
  \addtolength{\figwidth}{-\columnsep}
  \setlength{\figwidth}{0.5\figwidth}
  
  \begin{minipage}[t]{\figwidth}
    \mbox{}
    \vskip -7pt
    \centerline{\includegraphics[width=0.82\linewidth,angle=270]{fig5a.eps}}

  \end{minipage}
  \hfill
  \begin{minipage}[t]{\figwidth}
    \mbox{}
    \vskip -7pt
    \centerline{\includegraphics[width=0.82\linewidth,angle=270]{fig5b.eps}}

  \end{minipage}
\begin{minipage}[t]{\figwidth}
    \mbox{}
    \vskip -7pt
    \centerline{\includegraphics[width=1.1\linewidth,angle=0]{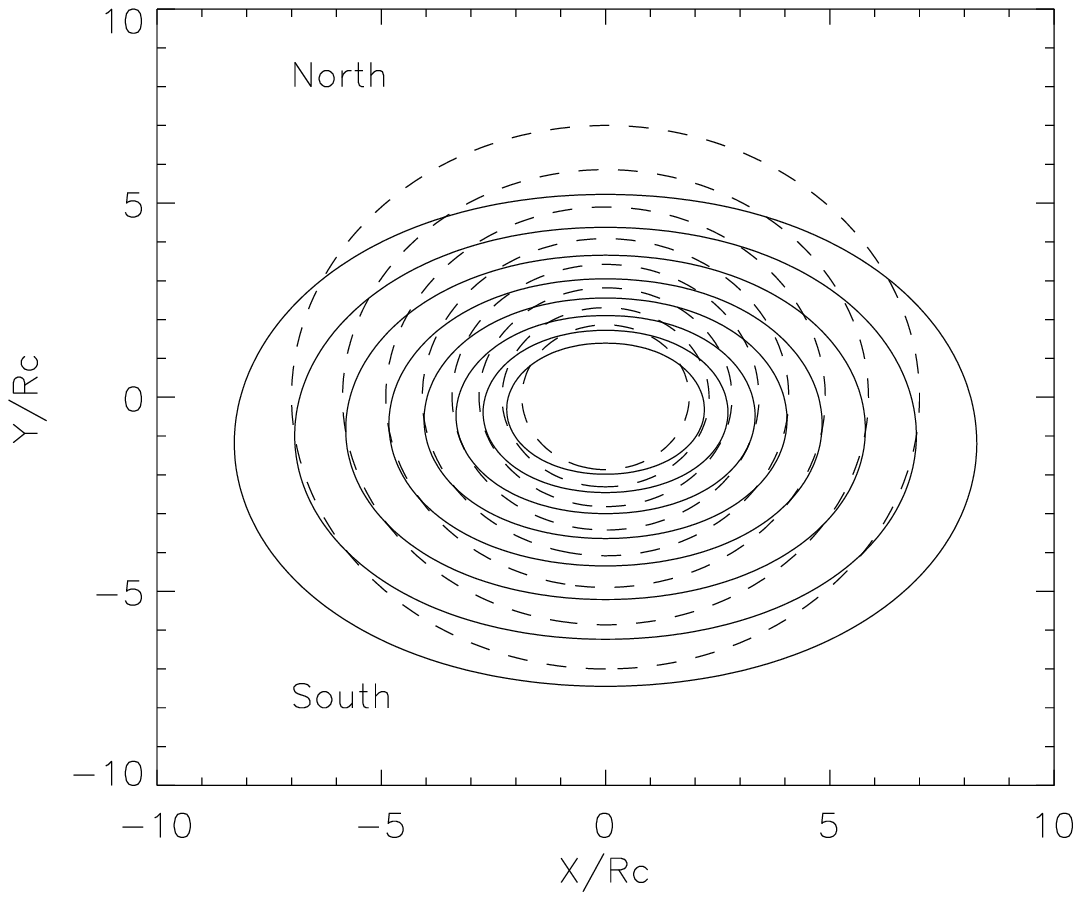}}

  \end{minipage}

    \caption{Top left: Half thickness of neutral hydrogen gas in Milky Way, {\bf model B}. The solid line is the model fitted in the North and dashed line in the South. Beyond $\sim$ 16 kpc the fit is quite good. The model parameters are given in Table 1. Top right: Rotation curve of the Galaxy for {\bf model B}. Bottom: Density contours of the dark matter halo of {\bf model B}. dashed lines are for the axisymmetric perfect ellipsoidal ($p=2$) halo. Solid lines are for the asymmetric lopsided $p=2$ dark matter halo with the second harmonic ($m=2$)$\sim$ 50\% of the first ($m=1$) one (see table 1). $R_c$ is the core radius of the dark matter halo. Contour levels are 0.05$\times \rho_{0}$ for the inner most one and decreasing by a factor of 2 outwardly. $\sigma_{H {\footnotesize I}}$ used for this model is 9.0 kms$^{-1}$. }
\end{figure*}

\noindent{\bf {Model B}}

Here we consider the velocity dispersion of H {\footnotesize I}, $\sigma_{H {\footnotesize I}} = 9$ km s$^{-1}$ slightly higher compared to model A and again flat out to 30 kpc. The increased velocity dispersion improves the fit by a very small amount below 16 kpc. With the same halo parameters as in model A, the fit is not good beyond 16 kpc. So we reduced the strength of the second harmonic ($\epsilon_{h}^{2}$) and when the second harmonic is decreased to $\sim 2/3$ of the first harmonic ($\epsilon_{h}^{1}$) we again recovered a good fit to the data. The resulting model halo with parameters $\epsilon_{h}^{1} =0.2$ and $\epsilon_{h}^{2}=0.14$ reproduced the N-S asymmetry in the Galaxy quite well (see Fig.~5). The rotation curves for the model in the two halves of the Galaxy are shown in the right panel of Fig.~5. At 25 kpc, the rotation speed in the north is $\sim 191$ km$^{-1}$ and in the south it is $\sim 212$ km$^{-1}$. It appears that the rotation curve in the south is more closer to the flat rotation curve (with $V_c = 220$ km$^{-1}$). The density contours of the model dark matter halo are shown in the bottom panel of Fig.~5. Due to the presence of a small component of second harmonic, the contours are elongated by a small amount compared to model A in the East-West direction of the Galaxy. This dark matter halo can be considered as a dominantly lopsided halo. This model again confirmed our earlier findings from model A. 

\begin{figure}[h]
\includegraphics[angle=270,scale=0.60]{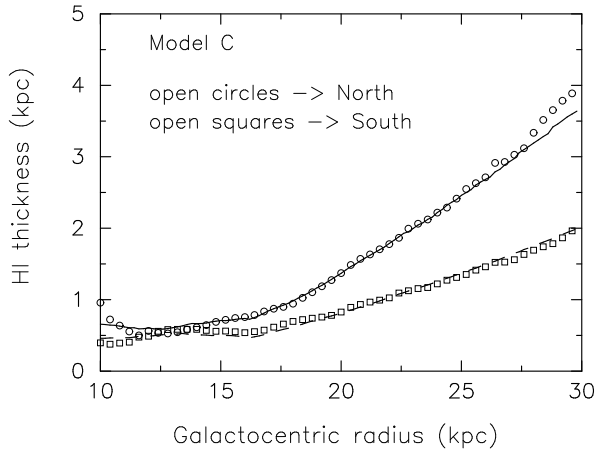}

\caption{Half thickness of neutral hydrogen gas in Milky Way, {\bf model C}. Solid line is the model fitted in the North and dashed line to the South. The H {\footnotesize I} velocity dispersion is quite high in the radial range 10--16 kpc. See Table 1 for the dark matter halo parameters.}
\end{figure}

\noindent{\bf {Model C}}

We constructed this model primarily to see if we can explain the observation also in the region $10 < R < 16$ kpc. We found that the mid-plane density distributions of H {\footnotesize I} and stars are almost same in the radial range $10 \le R \le 16$ kpc (see Fig.~8) and interestingly enough the half-thickness of H {\footnotesize I} in this radial range is almost the same as that of stars. Another important fact to be noted is that the H {\footnotesize I} half-thickness is roughly constant in this radial range ($10 \le R \le 16$ kpc). Now, we know that the thickness of H {\footnotesize I} $\sim \sigma_{H {\footnotesize I}}/\sqrt(G \rho_{mid})$, where $\rho_{mid}$ is the total mid-plane volume density. So we vary the H {\footnotesize I} velocity dispersion in the following manner in this radial range:  
\begin{eqnarray}
 \sigma_{H {\footnotesize I}} & = & 55.6 e^{-R/{2 R_{\sigma}}}   \mbox{kms$^{-1}$}    \mbox{ for $10 \le R \le 16~{\rm kpc}$ }.\\
                              & = & 8.5       \mbox{kms$^{-1}$}     \mbox{ $R \ge 16$ kpc} \nonumber        
\end{eqnarray}

\noindent From 16 kpc ($\sim$ 5 R$_{d}$) onwards $\sigma_{H {\footnotesize I}}$=8.5 km s$^{-1}$ and stays flat out to 30 kpc. We consider $R_{\sigma}=4370$ pc because a similar value in the case of stellar disk can produce a flat thickness (Lewis \& Freeman 1989) in the Galaxy. 

With the H {\footnotesize I} velocity dispersion varied in the above fashion (eq.[21]) and $\epsilon_{h}^{1}=0.2$ and $\epsilon_{h}^{2}$=0.18 (same as model A), we got an excellent fit to the observation (see Fig.~6). Model C provides the best fit to the observed data and reproduces the observed asymmetry in the thickness map of H {\footnotesize I}. In terms of halo asymmetry parameters, model C is exactly equal to the model A and so are the rotation curves. The density contours are the same as in Fig.~4. So model C confirms that our best fit dark matter halo model and again it is a lopsided dark matter halo with some elliptical perturbation.

\section{Comparison of different models}

\subsection{Measurement of asymmetry}
Given the radial variation of the H {\footnotesize I} thickness in different azimuths ($\phi$), we can quantify the degree of observed asymmetry by $\eta$ (see eq.[18]) in the two halves of the Galaxy. We used this parameter $\eta$ as a measure of the underlying asymmetry in the thickness distribution and to differentiate between the above models which seem to reproduce the observation quite well. The value of $\eta_{obs}$ is computed by assuming $\varphi_{k} = 90\,^{\circ}$ as discussed at the end of \S 3. Just by comparing the observed asymmetry ($\eta_{obs}$) and the model predicted asymmetry ($\eta_{model}$), we find that model p2L is the best match to the observation (see Table 1) but the velocity dispersion used in this model is quite high. Our next best model is model C based on the comparison of asymmetry and this model also has a very high velocity dispersion between 10 - 16 kpc. On the other hand model A is probably more reasonable because the H {\footnotesize I} velocity dispersion is not very high. The H {\footnotesize I} dispersion of model A is almost the same as that used recently by Kalberla et al. (2007); their solution is close to a single component model like we have considered with an effective constant velocity dispersion of 8.3 km s$^{-1}$. In all the four models of $p=2$ dark matter halo one thing is common and it is the parameter $\epsilon_{h}^{1} \sim 0.2$, the degree of lopsidedness. {\it So the different models are essentially built up around the same basic configuration, a lopsided halo}. To make the picture more clear, these models, being a function of two apparently independent parameters ($\sigma_{H {\footnotesize I}}$ and $\epsilon_{h}^{2}$), can be thought of as a family of models of lopsided dark matter halos. From Table 1, amongst the first three models, it is clear that as $\epsilon_{h}^{2}$ increases, $\sigma_{H {\footnotesize I}}$ decreases to make the fit better. One interesting fact about these family of models is that a better fit to the observed data demands that the {\it second harmonic be out of phase with the first one}.      

\begin{table}[h]
\caption[ ]{Asymmetry and the models of the $p=2$ lopsided dark matter halos}
\begin{flushleft}
\begin{tabular}{ccccccccccc}  \hline 
Model & $\sigma_{H {\footnotesize I}}$ &$\rho_{0}$ & $R_c$ & q & $\epsilon_{h}^{1}$ & $\epsilon_{h}^{2}$ & $\Sigma_{1.1}$  & $\eta_{obs}$  & $\eta_{model}$\\
        & (km s$^{-1})$ & M$_{\odot}$pc$^{-3}$ & kpc &  &        &    & M$_{\odot}$pc$^{-2}$  &  &  &          \\
\hline
p2L & 11.0 & 0.035 & 9.4 & 0.95 & 0.17 & 0.0 & 72.0 &0.262 & 0.261\\

A  &8.5 & 0.035 & 9.4 & 0.95 & 0.20 & 0.18 & 80.0 &0.262 & 0.265\\

B  &9.0 & 0.035 &9.4 &0.95& 0.20 & 0.14 & 78.0 &0.262 & 0.271\\

C & Eq.[21] & 0.035 &9.4 &0.95& 0.20 & 0.18 & 81.0 &0.262 & 0.260 \\
\hline
\end{tabular}
\end{flushleft}
\label{param}
\end{table}

\noindent In Table 1 $\eta_{obs}$ is obtained from the observed average  thickness of H {\footnotesize I} on both the halves. $\eta_{model}$ is from the fitted curves.

\subsection{Constraints from the surface mass density at Solar radius}
Apart from the constraint on the rotation curve due to Kerr \& Lynden-Bell (1986), we provide here our estimations of total surface density at the solar radius for various models and compare them with other measurements from the literature. The value of the total surface density at the solar radius put a strong and important constraint on any mass model of our Galaxy. Based on K dwarfs, Kuijken \& Gilmore (1991) provide us the total surface density within 1.1 kpc from the Galactic midplane $\Sigma_{1.1} = 71 \pm 6$ M$_{\odot}$pc$^{-2}$. By analysing HIPPARCOS K giants, Holmberg \& Flynn (2004) measure $\Sigma_{1.1} = 74 \pm 6$ M$_{\odot}$pc$^{-2}$. By modeling the H {\footnotesize I} thickness in the Galaxy, Kalberla et al. (2007) estimate $\Sigma_{1.1} = 79 \pm 2$ M$_{\odot}$pc$^{-2}$. Our best fit models are in good agreement with all the previous measurements. For example, for the model p2L $\Sigma_{1.1} = 72$ M$_{\odot}$pc$^{-2}$; for the model A, $\Sigma_{1.1} = 80$ M$_{\odot}$pc$^{-2}$; for the model B, $\Sigma_{1.1} = 78$ M$_{\odot}$pc$^{-2}$ and for model C, $\Sigma_{1.1} = 81$ M$_{\odot}$pc$^{-2}$. We used the total baryon mass surface density at the solar radius to be $\Sigma_{star +gas} = 50$ M$_{\odot}$pc$^{-2}$ which is within the limit ($48 \pm 8$ M$_{\odot}$pc$^{-2}$ at R$_{\odot}$) provided by Kuijken \& Gilmore (1989). Within $800$ pc from the Galactic plane Holmberg \& Flynn (2004) reports a surface density of $\Sigma_{0.8} = 65 \pm 6 $ M$_{\odot}$pc$^{-2}$ and from Kalberla et al. (2007) $\Sigma_{0.8} = 66.9 \pm 2 $ M$_{\odot}$pc$^{-2}$. In this context, our model p2L gives $\Sigma_{0.8} = 65 $ M$_{\odot}$pc$^{-2}$ and from model A, we get $\Sigma_{0.8} = 71 $ M$_{\odot}$pc$^{-2}$. In Fig.~7, we plot the total surface mass density (derived from modeling of the vertical hydrostatic equilibrium) of the Galaxy (including stars, gas and dark matter) within 1.1 kpc from the Galactic midplane as function of radius. It shows clearly that the surface density is higher in the Southern part of the Galaxy. In any case, our results on the surface density measurements in the solar neighbourhood are in close agreement with previously quoted values or within the quoted error bars.

\begin{figure}[h]
\includegraphics[angle=270,scale=0.30]{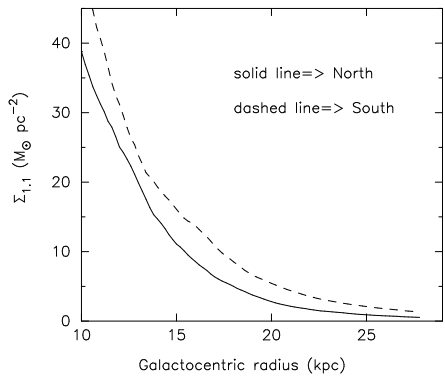} 
\caption{The total surface mass density in the Galaxy within 1.1 kpc from the Galactic midplane. It includes the stars, H {\footnotesize I} and the dark matter distribution from our model p2L. At 2R$_{\odot}$, the difference in the total surface densities between the North and South is $\sim 4 M_{\odot} pc^{-2}$. }
\end{figure}

\subsection {Total mass of the Galaxy}
Starting from the cosmological point view to studying galactic dynamics and especially in Galactic astronomy it is very important to have a good knowledge of the total mass of Milky Way. One of the complete analysis using most of the usual methods e.g. satellite kinematics, very high velocity stars, Local Group and Galactic rotation curve, Kochanek (1996) estimated the total mass of the Galaxy to be $\sim 4.9 \pm 1.1 \times 10^{11}$ M$_\odot$ within 50 kpc from the center of the Galaxy. Previous to this study, Lin, Jones \& Klemola (1995) found the mass of the Milky Way to be $5.5 \pm 1.0 \times 10^{11}$ M$_\odot$ within 100 kpc from the centre. By modeling the rotation curve, Dehnen \& Binney (1998) provided the Milky Way's mass within 100 kpc to be $7 \pm 2.5 \times 10^{11}$ M$_\odot$. From the $uvby - \beta$ survey of high velocity stars, Garcia Cole et al. (1999) suggest the mass of Milky Way to be $3.2 - 3.6 \times 10^{11}$ M$_\odot$ within 50 kpc from the Galactic centre. By using the kinematic information for Galactic satellites and halo objects, Sakamoto et al. (2003) estimated the mass of the Galaxy within 50 kpc (excluding Leo I) to be $5.0 - 5.5 \times 10^{11}$ M$_\odot$. In contrast to these studies, Kalberla et al. (2007) based on the H {\footnotesize I} thickness modeling in the Galaxy found the total mass of the Milky Way disk itself to be $2.9 \pm 0.1 \times 10^{11}$ M$_\odot$. The most recent analysis based on the Blue Horizontal-Branch halo stars from the SDSS, Xue et al. (2008) estimate the mass of our Galaxy to be $4.0 \pm 0.7 \times 10^{11}$ M$_\odot$ within 60 kpc from the Galactic centre. Based on the modeling of H {\footnotesize I} thickness, we find the total mass of Milky Way to be $\sim 3.3 \times 10^{11}$ M$_\odot$ within 100 kpc from the center and within 50 kpc it is about $3.1 \times 10^{11}$ M$_\odot$. The mass of stellar disk is $\sim 4.2 \times 10^{10}$ M$_\odot$ and the halo mass within 100 kpc is $\sim 2.53 \times 10^{11}$ M$_\odot$. The difference between the halo masses in the North and South is $\sim 2 - 3 \times 10^{10}$ M$_\odot$ in our models of asymmetric halos. It is true that the total mass of the Galaxy saturates beyond about 100 kpc because the density profile of our p=2 dark matter falls faster ($\propto R^{-4}$ at $R >> R_c$) than the the isothermal one. But within 100 kpc, our estimates are in good agreement with most of the previous mass estimates of Milky Way.

\section{Comparison with Previous Work}
The most recent work which has extensively used the LAB survey H {\footnotesize I} data are  Levine et al (2006a) and Kalberla et al (2007). Of these two, Kalberla et al. (2007) has studied in considerable detail the behaviour of the H {\footnotesize I} thickness in the Galaxy and found that in order to explain the observation they needed beside a massive dark matter halo, a dark matter disk and a dark matter ring. Such configurations of the dark matter for our Galaxy is troublesome for the CDM paradigm. 

\noindent Here, we would like to point out few similarities and differences of our method with the previous ones. First of all, one of our best-fit models, namely model A, uses a constant H {\footnotesize I} velocity dispersion of 8.5 km s$^{-1}$ which is close to what Kalberla et al. (2007) used for the their model namely 8.3 km s$^{-1}$. To make it more clear, although we do not use explicitly the different phases of ISM (e.g. CNM and WNM) to model the H {\footnotesize I} thickness map, the state of the ISM is roughly the same in our model as it is in Kalberla et al.~(2007). In contrast to Kalberla et al. (2007), we consider non-axisymmetric models of dark matter halo and also a mild lopsided stellar disk. From Fig. 8, it is clear that the gas self-gravity is comparable to the stellar one and we take into account the full self-gravity of the H {\footnotesize I} gas in deriving the thickness map of the gas. 
 
There is a marked difference in the averaged flaring curves in the North and the South derived in our paper with that in Kalberla et al. ~(2007) which excludes a region $90\,^{\circ} < \varphi < 110\,^{\circ}$ in the North showing very high flaring in the H {\footnotesize I} thickness map. On the other hand, if this region is included in the averaging process, the averaged observed gas thickness would increase by $\sim 20\%$ and if the H {\footnotesize I} velocity dispersion remains constant, one would expect a less massive dark matter halo from a simple minded calculation. Similarly, if we exclude this region the thickness in the North reduces to $\sim 3.4$ kpc.

\noindent Another important fact about the LAB survey data is the pronounced N-S asymmetry in the gas thickness distribution. Certainly, an axisymmetric model of the Galaxy can not reproduce such asymmetry in the data. In this context, recent work by Sanchez-Salcedo et al. (2008) have also shown that MOND provides a reasonably good fit to the azimuthaly averaged flaring in the H {\footnotesize I} gas using the same LAB survey data. Now for an axisymmetric baryon distribution, the MOND potential would also be axisymmetric, because the differential operator acting on the potential in the Poisson equation is rotationally invariant. Hence, it would probably be very hard to reproduce the observed N-S asymmetry in the H {\footnotesize I} thickness map. Whereas our family of lopsided dark matter halos naturally explain the observed North-South asymmetry. 

\noindent  The density distribution of the dark matter halo (namely the $p=2$) in our model is similar to that obtained by Narayan et al. (2005) (namely a $p=2$ halo as the best fit model) based on the Wouterloot et al. (1990) H {\footnotesize I} flaring data. The axisymmetric version of our model is similar to that used by Narayan et al. (2005). 
   
\begin{figure}[h]
\includegraphics[angle=270,scale=0.30]{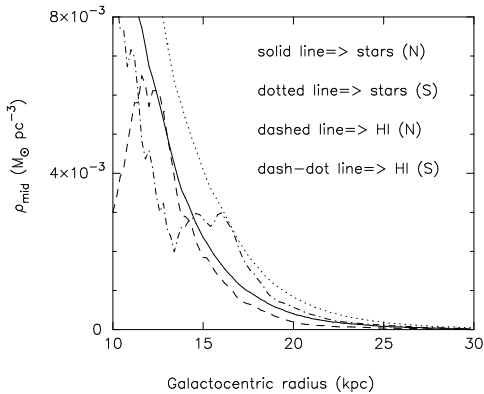}
\caption{Mid-plane volume density of stars and gas in the two halves of the Galaxy. Density distributions are shown for the model A. For other models they are almost the same. Note that H {\footnotesize I} gas is comparable to the stars. N= North and S= South}
\end{figure}

\section{Discussion}
\noindent {\bf (i) Self-gravity of the atomic hydrogen gas}

\noindent In the radial range from 10 - 16 kpc, we find that azimuthally averaged mid-plane volume density of the atomic hydrogen gas is comparable to that of the stars (see Fig.~8) and beyond this radial range gas is in fact dominating over the stars. This suggests that the self-gravity of the H {\footnotesize I} gas is quite important in determining its thickness and thereby the nature of dark matter halo. By self-gravity of the gas we mean that the gas is held by its own gravity. In other words, we include the gas density in the Poisson equation to derive its contribution to the total potential of the Galaxy. Without any self-gravity, the gas would move like a test particle under the potential of the Galaxy. So it would be interesting to check the change brought about by excluding the H {\footnotesize I} self-gravity on its vertical scale-height. The difference is not negligible, it is seen to be about 10-15$\%$ within the optical disk ($R< 4R_d$). This could be because these regions are dominated by the stellar disk and by the dark matter respectively. For the region 4 $< R/R_d <$ 6, the difference is substantial ($\sim$20 - 30$\%$) suggesting that in this range, the gas gravity is very important in negotiating the hydrostatic equilibrium for the H {\footnotesize I} layer. Thus neglecting it may lead to a serious overestimate of the H {\footnotesize I} scaleheight in general at all radii in the outer Galaxy and to explain the observed gas scale-height one may need to invoke a heavier dark matter halo.

\noindent {\bf (ii) Molecular hydrogen gas}

In the present work, we are mostly concerned with the very outer region of the Galactic disk especially beyond 16 kpc. The particular reason for this is that the thickness curve of the atomic hydrogen gas in the northern hemisphere of the Galaxy deviates noticeably from the southern one beyond this region. The molecular hydrogen gas extends upto about 17 kpc in the Galaxy (Wouterloot et al.~1990) and beyond around this region there is little data. Since our calculation of gas thickness is local (in the sense that we do not consider the gas gravity in global sense) we have neglected the effect of molecular hydrogen on the H {\footnotesize I} thickness throughout our work. In fact, the absence of molecular hydrogen gas beyond about 16 kpc makes it more convenient in disentangling the effect of dark matter halo on the H {\footnotesize I} gas.  

\noindent {\bf (iii) Effect of Galactic constants}

We have built the stellar disk model based on the IAU-recommended values for the galactic constants - $R_{\circ}$ = 8.5 kpc and $\Theta_{\circ}$ = 220 km s$^{-1}$. It would definitely be interesting and also worthwhile to know how the results for the halo density profile vary with the assumed galactic constants. For example, Olling \& Merrifield (2001) find the effect of varying these constants on the inferred axis ratio of the halo. Unfortunately, all the observational inputs for our model, like the H {\footnotesize I} and H$_2$ surface densities, H {\footnotesize I} scale-height and the stellar velocity dispersion, are based on the IAU-recommended galactic constants and rescaling them for different values of the constants is beyond the scope of this paper.

\noindent {\bf (iv) Rotation curve and gas thickness data}

Different rotation curves produce different degrees of flaring because of
the dependence of R on $\Theta$ through equation 1 of Levine et al. (2006a).
The rotation curves used in the present analysis differ by a small amount
$<$ 10\% from the flat rotation curve, $\Theta (R)$ = 220 km s$^{-1}$, used
by Levine et al. (2006a).  An error analysis shows that R changes
approximately by $\sim \Theta/\Theta_{\odot}$, where $\Theta_{\odot}$ is the
circular speed at the solar position. Thus, $dR/d\Theta$, the sensitivity of
an error in the distance to an error in the circular speed produces an error
in the distance of a parcel of gas from the center of about $10\%$ at 30
kpc.   Because all values of Galactic latitude are small at large R, this
translates to an error in the thickness of the gas layer by no more that
10\%, and then only at the largest distances.

Although the differences in the rotation curves used by us and by Levine et
al. (2006a) implies that the two analyses are not exactly commensurate, we
are trying to explain a factor of 2 increase in the thickness of the gas
layer of the Milky Way from one hemisphere to the other, an order of
magnitude larger than the maximum 10\% effect caused by differences in the
rotation curves. Note that the factor of 2 change in the scale height occurs
at all radii beyond about R = 17 kpc and the effect on $\Delta R$ will be
smaller at smaller radii because $\Delta \Theta$ is also smaller.
Therefore, while there is a 10\% effect at the largest distances, the most
it will do is to have a 10\% effect on the scaling of the model.

It is worth mentioning at this point that we do not solve the vertical
hydrostatic equilibrium considering epicylic orbit correction for the gas.
For this one needs to solve the Jeans equation and Poisson equation
self-consistently, which is considerably more complex than what we do here,
and in any event probably results in only a small correction.  Our primary
aim is to understand the nature of the observed asymmetry and build a first
order model to explain it.

\noindent {\bf (v) Effect of Galactic warp}

The vertical structure of the Milky Way's disk is fairly complicated. The warp in the H {\footnotesize I} disk is very asymmetric; in the northern hemisphere the disk mid-plane rises to a height of about 4 kpc, while in the south it goes down to about 1 kpc and then again comes back to the undisturbed mid-plane (Kerr 1957;  Burton 1988; Levine et al. 2006a). Since the warp is a global feature in the Galaxy, it will have its own self-gravity to affect the vertical oscillations in the disk. The effective vertical oscillation frequency in the disk can be written as $\nu_{eff} = \sqrt{\nu_{d}^2 + \nu_{warp}^2}$, where $\nu_{d}$ is the vertical frequency of the unperturbed disk and $\nu_{warp}$ is the extra vertical frequency due to the enhanced self-gravity of the warp. Because the warp is global in nature, $\nu_{warp}$ is an integral quantity (see eq.[4b] in Saha \& Jog 2006). However, an analytic form for $\nu_{warp}$ can be written in a local sense via the WKB analysis and it is $\sim \sqrt{2\pi G \Sigma_{d}|k|}$ (Binney \& Tremaine 1987), where $|k|$ is the wavenumber for the warp. Since we are solving the vertical hydrostatic equilibrium locally, the contribution from the global warp ($|k| \longrightarrow 0$) becomes insignificant. In order to account for the warp in the gas thickness, one has to formulate the vertical hydrostatic equilibrium as an integral problem which we plan for the future.  

\noindent {\bf (vi) Galactic spiral structure}

The recent work by Levine et al. (2006b), based on the 21 cm LAB Galactic H {\footnotesize I} survey, reveals a multi-armed spiral structure of our Galaxy. Their study shows a good correlation between the positions of the spiral arms and the H {\footnotesize I} thickness. This is an expected behaviour because the presence of the spiral arms would cause enhanced gravity and scattering of the H {\footnotesize I} clouds would not change the vertical velocity dispersion appreciably, leading to a decrement in the H {\footnotesize I} thickness. However, we do not expect the thickness of H {\footnotesize I} to change appreciably because the strength of the perturbed surface density does not vary strongly as a function of the Galactocentric radius along an arm or even from arm to arm (Levine et al. 2006b). In the present paper, under the zeroth order approximation, we worked with the average data in the northern and southern halves of the Galaxy respectively. This is certainly an incomplete modeling of the data and in future we expect to construct a self-consistent formulation of the problem to include the spiral structure of the Galaxy.

\noindent {\bf (vii) Uniqueness of the Lopsided halo model}

In astronomy, it is quite a hard job to prove directly the uniqueness of a proposed model which fits a given set of observational data well. One obvious way of approach is to compare different theoretical models against the given data set. Given the various uncertainties in the observation and our incomplete understanding of the physics of our Galaxy, such a comparison may even lead to degeneracies and bayesian analysis would probably be the appropriate way out to lifting such degeneracies or discard other obvious models based on physical grounds. Before we go into discussing other possible theoretical models, we would like to remind the reader some basic facts just for a recap. We are dealing with the outer region of the Galaxy and especially beyond $R > 16$ kpc which is $\sim 5$ disk scale length and this is also the typical size of the stellar disk in a disk galaxy. So the influence of the stellar disk on the vertical distribution of the gas is almost negligible compared to that due to the dark matter halo. 

Certainly, the internal disk instabilities alone (being normally weak, except the bar instability) are unlikely to produce the observed asymmetry in the gas thickness distribution. 

One of the possible alternative models is an off-centered dark matter halo with respect to the Milky Way's disk. Such a model of an off-centered axisymmetric halo has been used previously by Levine \& Sparke (1998) to generate lopsidedness in the disk. As pointed out by these authors that this method is effective when the disk lies within the core radius (almost constant density region of the halo) of the halo and as result is more efficient in dwarf galaxies rather than in luminous galaxies. In this context, one of our halo model namely $p=2$ halo whose core radius is $\sim 9$ kpc as compared to the disk scale lenth $\sim 3$ kpc could have been a possible case for investigation. However, note that the N-S asymmetry in the gas distribution begins beyond $16$ kpc which is roughly $\sim 2 \times R_c$ of the halo and at this radius, the disk is no more in a nearly constant density region of the halo, making it harder to maintain the lopsidedness. In any case, this is an interesting possibility to be investigated in a future problem. 

Other possible model is an asymmetric gas accretion onto the disk as proposed by Bournaud et al. (2005) to reproduce lopsidedness observed in the galactic disk. To produce strong lopsidedness as observed, one needs the accretion of the cold gas through the cosmological filaments to be highly asymmetric and in reality it is not clear if the gas accretion is asymmetric to such a degree. On the other hand, it makes sense to think in this direction because the H {\footnotesize I} distribution looks more disturbed in the North rather than in the South. At this point, it is worthwhile to mention that the gas actually contributes very little to the rotation curve (mostly dominated by the dark matter); so an axisymmetric $p=1$ halo with an asymmetric gas accretion again may not be the good candidate for the present observation. However, an axisymmetric $p=2$ halo with an asymmetric gas accretion could have been a possible candidate, but its beyond the scope of the present paper to examine such a model in considerable detail. 

On the other hand, our lopsided halo models are more natural to occur in cosmological scenario. Tidal interactions or large scale tidal harassment or major mergers of neighbouring dark matter halos are more likely to produce large scale perturbations in the dark matter halo. And since the CDM halos are collisionless object the survival of such global perturbations in the halo is less of a problem.     

\section{Conclusions}

We have analyzed both axisymmetric and non-axisymmetric configurations of dark matter halos with a consistent picture of the ISM to explain the observed nature of the North-South asymmetry in the thickness distribution of the H {\footnotesize I} gas. Below we summarize the main results:
 
\noindent With a model of the ISM that has reasonable values of the gas velocity dispersion, an isothermal dark matter halo producing a flat rotation curve with $V_c$=220 km s$^{-1}$ cannot produce the observed flaring in the H {\footnotesize I} gas thickness.

\noindent We show that the nature of the systematic North-South asymmetry in the H {\footnotesize I} thickness map is gravitational in nature. An axisymmetric dark matter halo with different values of the  H {\footnotesize I} velocity dispersion in the two halves of the Galaxy can not reproduce this asymmetry. The observed asymmetry in the thickness map of neutral hydrogen gas is apparently not the result of purely gas dynamical effects. 

\noindent We show that a purely $p=1.5$ or $p=2$ lopsided dark matter halo also cannot explain the observed North-South asymmetry. For a purely $p=2$ lopsided halo, the H {\footnotesize I} velocity dispersion has to be unreasonably large to come close to the observation and even then, the fit is not very good.

\noindent Finally, we come up with a configuration of the dark matter halo in which some amount of second harmonic ($m=2$) is superposed out of phase onto a purely $p=2$ lopsided halo. For the best fit models (A \& C), the values of lopsidedness and elliptical perturbation are $\epsilon_{h}^{1}=0.2$ and $\epsilon_{h}^{2}=0.18$ respectively. We call such a halo  an elliptically perturbed lopsided dark matter halo which can explain the observed North-South asymmetry. Basically, the emerging picture of the dark matter halo of the Milky Way is dominantly lopsided in nature. In such a halo, the density falls off faster than the $p=1$ isothermal dark matter halo. The emerging model of the asymmetric dark matter halo is supported by the halos formed in the recent cosmological N-body simulation.

\noindent{\bf Acknowledgment:} 
 We thank the referee for the critical and constructive comments which has improved the contents of the paper substantially. KS would like to thank the Raman Research Institute for supporting a visitorship during which an initial draft of the paper started. LB and EL would like to acknowledge support of NSF grant AST-0540567 to the University of California.

\end{document}